\documentclass[aps,twocolumn,pra,showpacs,floatfix]{revtex4}
\usepackage{epsfig}
\usepackage{graphicx}
\usepackage{dcolumn}
\usepackage{amsmath}

\begin{document}


\title{Relativistic corrections to transition frequencies of Fe~I and
search for variation of the fine structure constant}

\author{V. A. Dzuba}
\affiliation{School of Physics, University of New South Wales,
Sydney 2052, Australia}
\author{V. V. Flambaum}
\affiliation{School of Physics, University of New South Wales,
Sydney 2052, Australia}

\date{\today}

\begin{abstract}

Relativistic energy shifts of the low energy levels of Fe 
have been calculated using the Dirac-Hartree-Fock and 
configuration interaction techniques. The results are to be used in
the search for the space-time variation of the fine structure constant
in quasar absorption spectra. The values of the shifts are the largest 
among those used in the analysis so far. This makes Fe a good candidate
for the inclusion into the analysis. 

\end{abstract}


\maketitle

\section{Introduction}

Theories unifying gravity with other interactions suggest a possibility
of temporal and spatial variations of the fundamental  constants of nature;
reviews of these theories and results of measurement can be found
in Refs.~\cite{Uzan,Flambaum07a}. Strong evidence that the fine-structure 
constant might be smaller about ten billion years ago was found in the 
analysis of quasar absorption 
spectra~\cite{Webb99,Webb01,Murphy01a,Murphy01b,Murphy01c,Murphy01d}.
This result was obtained using the data from the Keck
telescope in Hawaii. However, an analysis of the data
from the VLT telescope in Chile, performed by different groups~\cite{vlt1,vlt2}
gave a null result. There is an outgoing debate
in the literature about possible reasons for the disagreement 
(see. e.g.~\cite{Flambaum07,Srianand07}).

All these results were obtained with the use of the so called
many-multiplet method suggested in Ref.~\cite{Dzuba99}.
This method requires calculation of  relativistic corrections
to frequencies of atomic transitions to reveal their dependence
on the fine-structure constant. Atomic calculations for a large
number of transitions in many atoms and ions of astrophysical 
interest were reported in Refs.~\cite{Dzuba99a,Dzuba01,Dzuba02,
Berengut04,Berengut05,Berengut06,archDzuba,Porsev,Johnson,Savukov}.
In present paper we perform similar calculations for neutral iron 
which was never considered before. Until very recent we were unaware 
about any lines of neutral iron observed in the quasar absorption 
spectra. It was Prof. P. Molaro~\cite{Molaro} who brought to our 
attention the existence of such data and the intention of his group 
to use them in the analysis.

Using Fe~I in the search for variation of the fine structure constant
has several advantages. First, the values of the relativistic energy
shifts are large due to relatively larger nuclear charge ($Z$=26)
and strong configuration mixing. Second, these values vary strongly
from state to state which makes it hard to mimic the effect of varying
fine structure constant by any systematics. Finally, iron is very 
abundant element in the universe. Great part of the previous analysis
was performed using the data from Fe~II. 

Calculations for Fe~I are difficult due to large number of valence electrons
and strong configuration mixing. Its ground state configuration for
outermost electrons is $3d^64s^2$ which is strongly mixed with the
$3d^74s$ configuration. There is strong configuration mixing for the 
$3d^64s4p$ and $3d^74p$ odd-parity configurations for the excited
states. With eight valence electrons and strong configuration mixing
full-scale accurate {\em ab initio} calculations for Fe~I would require
enormous computer power. We have chosen a different approach. Below
we report a simple method which is specially designed for systems
with strong mixing of several distinct configurations. It combines 
{\em ab initio} Hartree-Fock and configuration interaction (CI)
techniques with some semi-empirical fitting and gives very
reasonable results at very low cost in terms of computer power.
The approach is similar to the well-known multi-configuration
relativistic Hartree-Fock method (see, e.g.~\cite{Grant}) and
can probably be considered as a simple version of it.
The accuracy for the energy levels of Fe~I is within few per cent
of experimental values while estimated accuracy for the relativistic
energy shifts is on the level of 20 to 30\%. 
Due to strong configuration mixing the results are sensitive to the 
distances between energy levels. Therefore, special care has been
taken to reproduce experimental positions of the energy levels.

\section{Method}

It is convenient to present the dependence of atomic frequencies on
the fine-structure constant $\alpha$ in the vicinity of its physical
value $\alpha_0$ in the form
\begin{equation}
  \omega(x) = \omega_0 + qx,
\label{omega}
\end{equation}
where $\omega_0$ is the laboratory value of the frequency and
$x = (\alpha/\alpha_0)^2-1$, $q$ is the coefficient which is to be
found from atomic calculations. Note that
\begin{equation}
 q = \left .\frac{d\omega}{dx}\right|_{x=0}.
\label{qq}
\end{equation}
To calculate this derivative numerically we use
\begin{equation}
  q \approx  \frac{\omega(+\delta) - \omega(-\delta)}{2\delta}.
\label{deriv}
\end{equation}
Here $\delta$ must be small to exclude non-linear in $\alpha^2$ terms.
In the present calculations we use $\delta = 0.05$, which leads to
\begin{equation}
  q \approx  10 \left(\omega(+0.05) - \omega(-0.05)\right).
\label{deriv05}
\end{equation}
To calculate the coefficients $q$ using (\ref{deriv05}),  $\alpha$ must be varied
in the computer code.  Therefore, it is convenient to use a form
of the single electron wave function in which the dependence on $\alpha$ is
explicitly shown (we use atomic units in which $e=\hbar=1, \alpha = 1/c$)
\begin{equation}
    \psi(r)_{njlm}=\frac{1}{r}\left(\begin {array}{c}
    f_{v}(r)\Omega(\mathbf{n})_{\mathit{jlm}}  \\[0.2ex]
    i\alpha g_{v}(r)  \widetilde{ \Omega}(\mathbf{n})_{\mathit{jlm}}
    \end{array} \right),
\label{psi}
\end{equation}
where $n$ is the principal quantum number and an index $v$
replaces the three-number set $n,j,l$.
This leads to a form of radial equation for single-electron
orbitals which also explicitly depends on $\alpha$:
\begin{equation}
    \begin {array}{c} \dfrac{df_v}{dr}+\dfrac{\kappa_{v}}{r}f_v(r)-
    \left[2+\alpha^{2}(\epsilon_{v}-\hat{V}_{HF})\right]g_v(r)=0,  \\[0.5ex]
    \dfrac{dg_v}{dr}-\dfrac{\kappa_{v}}{r}f_v(r)+(\epsilon_{v}-
    \hat{V}_{HF})f_v(r)=0, \end{array}
\label{Dirac}
\end{equation}
here $\kappa=(-1)^{l+j+1/2}(j+1/2)$,
and $\hat{V}_{HF}$ is the Hartree-Fock potential.
Equation (\ref{Dirac}) with $\alpha = \alpha_0 \sqrt{\delta +1}$
and different Hartree-Fock potential $\hat{V}_{HF}$ for
different configurations is used to construct single-electron orbitals.

\begin{table}
\caption{Even and odd configurations of Fe and effective core
polarizability $\alpha_p$ (a.u.) used in the calculations.}
\label{sets}
\begin{ruledtabular}
  \begin{tabular}{l l l l}
\multicolumn{1}{c}{Set} &\multicolumn{1}{c}{Parity} & 
\multicolumn{1}{c}{Configuration} & 
\multicolumn{1}{c}{$\alpha_p$} \\
\hline
1 & Even &  $3d^64s^2$ & 0.4 \\
2 & Even &  $3d^74s$   & 0.4192 \\
3 & Even &  $3d^64p^2$ & 0.4 \\
4 & Even &  $3d^8$     & 0.465 \\
\hline
5 & Odd  &  $3d^64s4p$ & 0.39 \\
6 & Odd  &  $3d^74p$   & 0.412 \\
7 & Odd  &  $3d^54s^24p$ & 0.409 \\
\end{tabular}
\end{ruledtabular}
\end{table}

Table~\ref{sets} lists configurations considered in present work. First four
are even configurations and other three are odd configurations. 
We perform self-consistent Hartree-Fock calculations for each configuration
separately. This allows to account for the fact that single-electron 
states depend on the configurations. For example, the $3d$ state in the
$3d^64s^2$ configuration is not the same as the $3d$ state in the $3d^74s$
configuration. In principle, it is possible to account for these 
differences in the CI calculations. One would need to have a complete set
of single-electron states and construct many-electron basis states by
redistributing valence electrons over these single-electron basis states.
Then actual many-electron states are found by diagonalization of matrix
of the effective CI Hamiltonian. This approach works very well for the case 
of two or three valence electrons (see, e.g.~\cite{JETP,Kozlov96,Johnson98}).
However, for eight valence electrons it would lead to a matrix of enormous
size making it practically impossible to saturate the basis while using limited
computer power. The results with unsaturated basis are very unstable and
strongly depend on where the basis is truncated. Therefore, we prefer 
to account for the differences in the configurations on the Hartree-Fock
rather than CI stage of the calculations.

The self-consistent Hartree-Fock procedure is done for every configuration
listed in Table~\ref{sets} separately. Then valence states found in the 
Hartree-Fock calculations are used as basis states for the CI calculations.
It is important for the CI method that atomic core ($1s^2 \dots 3p^6$) remains
the same for all configurations. We use the core which corresponds to the
ground state configuration. Change in the core due to change of the valence
state is small and can be neglected. This is because core states are not
sensitive to the potential from the electrons which are on large distances
(like $4s$ and $4p$ electrons). The $3d$ electrons are on smaller distances
and have larger effect on atomic core. However, in most of the cases 
(see Table~\ref{sets}) only one among six $3d$ electrons change its state. 
Therefore their effect on atomic core is also small. More detailed
discussion on the effect of valence electrons on atomic core can be 
found in Refs.~\cite{VN,VN1}.

All configurations in Table~\ref{sets} correspond to an open-shell
system. We perform the calculations staying within central-field
approximation but using fractional occupation numbers. As a result
we have 23 singe-electron basis states for valence electrons:
$3d^{(i)}_{3/2},3d^{(i)}_{5/2},4s^{(i)},4p^{(i)}_{1/2},4p^{(i)}_{3/2}$.
Here index $i$ is the set number (as in Table~\ref{sets}).
Note that total number of basis states is less than 5 times number
of sets 
because many configurations don't include particular single-electron 
states. Note also that our basis set is non-orthogonal, e.g.
$0 < \langle 3d^{(i)}_{3/2}|3d^{(j)}_{3/2} \rangle <1$. The implications
of this fact will be discussed below.

The effective Hamiltonian for valence electrons has the form
\begin{equation}
  \hat H^{\rm eff} = \sum_{i=1}^8 \hat h_{1i} + 
  \sum_{i < j}^8 e^2/r_{ij},
\label{heff}
\end{equation}
$\hat h_1(r_i)$ is the one-electron part of the Hamiltonian
\begin{equation}
  \hat h_1 = c \mathbf{\alpha \cdot p} + (\beta -1)mc^2 - \frac{Ze^2}{r} 
 + V_{core} + \delta V.
\label{h1}
\end{equation}
Here $\mathbf{\alpha}$ and $\beta$ are Dirac matrixes, $V_{core}$ is
Hartree-Fock potential due to 18 core electrons ($1s^2 \dots 3p^6$) 
and $\delta V$
is the term which simulates the effect of the correlations between core
and valence electrons. It is often called {\em polarization potential} and
has the form
\begin{equation}
  \delta V = - \frac{\alpha_p}{2(r^4+a^4)}.
\label{dV}
\end{equation}
Here $\alpha_p$ is polarization of the core and $a$ is a cut-off parameter
(we use $a = a_B$).
The form of the $\delta V$ is chosen to coincide with the standard polarization
potential on large distances ($-\alpha_p/2r^4$). However we use it on distances
where valence electrons are localized. This distances are not large, especially
for the $3d$ electrons. Therefore we consider $\delta V$ as only rough 
approximation to real correlation interaction between core and valence
electrons and treat $\alpha_p$ as fitting parameters. The values of $\alpha_p$
for each configuration of interest are presented in Table~\ref{sets}.
They are chosen to fit the experimental position of the configurations
relative to each other. The value of $\alpha_p$ for the $3d^64p^2$ 
configuration is taken to be the same as for the ground state configuration  
because actual position of this configuration in the
energy spectrum is not known. For all configurations the values of
$\alpha_p$ are very close. This is not a surprise since the core is 
always the same. One can probably say that small difference in $\alpha_p$
for different configurations simulates the effect of incompleteness of the
basis and other imperfections in the calculations.

\subsection{CI calculations with a non-orthogonal basis.}

As it was mentioned above we have a set of single-electron states which is 
not orthogonal. The $3d$, $4s$ and $4p$ states in the configurations listed
in Table~\ref{sets} are similar but not the same. In principle, it may lead
to complication in the CI procedure, starting from non-orthogonality
of many-electron basis states which would lead in turn to complications in
calculation of matrix elements and matrix diagonalization.
However, most of these complications can be avoided by appropriate selection of
the configurations included in the calculations. It is sufficient to 
obey the two rules:
\begin{itemize}
\item Forbid configurations which have singe-electron states taken from different
sets, e.g. $3d^m_i3d^n_j4s_k4s_l$. Here $i,j,k$ and $l$ are set numbers as
in Table~\ref{sets} ($i \neq j$ or/and $k \neq l$)
and $m$ and $n$ are number of electrons in each of the
$3d$ state ($m+n=6$).
\item Don't generate additional configurations by exciting electrons
to the orbitals of the same symmetry, e.g. $3d^64s^2 \rightarrow 3d^64s5s$.
\end{itemize}
In present calculations we use only those configurations which are listed 
in Table~\ref{sets}.
 
If all single-electron states for every many-electron
basis state are taken from the same set then the many-electron basis states
remain orthogonal to each other. Indeed, states of the same configuration
are orthogonal to each other as in the standard CI technique. States of
different configurations are orthogonal because at least one electron
changes its angular symmetry in the transition between the configurations.
For example all states of the $3d^64s^2$ configuration are orthogonal to
all states of the  $3d^74s$ configuration because of the $s - d$ transition
involved.

Since many-electron basis functions remain orthogonal matrix diagonalization
is not affected. Calculation of the matrix elements between states of the same
configuration is not affected as well. The only part of the CI procedure 
which is affected is calculation of matrix elements between basis states
of different configurations. Here single electron part $\hat h_1$ (\ref{h1})
of the Hamiltonian does not contribute because this is a scalar operator
which cannot change angular symmetry of a single-electron state.
Only Coulomb integrals contribute to the matrix elements and
their calculation  must be accomplished by the product
of overlaps between similar states from different sets. For example,
Coulomb interaction between the  $3d^64s^2$ and $3d^64p^2$ configurations
has the form (in non-relativistic notations):
\[
F_1(4s_1,4p_3,4s_1,4p_3)\langle 3d_1|3d_3 \rangle^6.
\]
Here $F_1$ is dipole Coulomb integral, indexes 1 and 3 numerate basis sets
as in Table~\ref{sets}, $\langle 3d_1|3d_3 \rangle$ is the overlap between
different $3d$ functions.

\section{Results and discussion}

Neutral iron is an interesting system as a challenge for the calculations
and as a candidate for the search of the variation of the fine structure
constant. There is strong configuration mixing between the $3d^64s^2$ 
and the $3d^74s$ even configurations in the ground state and the
$3d^64s4p$ and the $3d^74p$ configurations for the odd excited states.
The latter mixing is a fortunate feature which makes Fe~I a convenient
object for the analysis. Let us elaborate. It is important to have
relativistic frequency shifts of the atomic transitions used in the analysis
to be as large as possible. The value of the shift depends on how many
electrons change their states in the transitions and how large is the
change of electron momentum in each single-electron transition
(see, e.g.~\cite{Dzuba99a}). The transition between $3d^64s^2$ and
$3d^64s4p$ configurations is basically a $4s - 4p$ transition.
However, mixing with the $3d^74s$ configuration in the upper state 
adds one more single-electron transition ($4s - 3d$) and makes the
frequency shift larger. Note that the presence of both the $3d^64s4p$
and the $3d^74p$ configurations is important. The first configuration 
is needed for the 
transition to the ground state to be strong electric dipole transition,
otherwise it will not be observed. The second configuration is needed for
the relativistic frequency shift to be large. It is fortunate that
strong configuration mixing between these two configurations takes 
place for most of the low odd states of Fe~I.

On the other hand this strong configuration mixing is a big challenge
for the calculations. It makes the results for the relativistic
energy shifts (the $q$-coefficients) to be unstable since they are
very sensitive to the value of the mixing. Note that the configuration
mixing in the ground state is also important. The admixture of the 
$3d^74s$ configuration adds the contribution of the $3d - 4p$
transition to the relativistic frequency shift. This contribution
has an opposite sing as compared to the $4s - 4p$ transition 
which adds to the instability of the results.

Since configuration mixing is very sensitive to the energy intervals
between the states the most reliable results can be obtained in the 
calculations which reproduce correctly experimental spectrum. 
In present calculations this is achieved with the use of the 
core polarization term (\ref{dV}) in the Hamiltonian and fitting
the data by changing the core polarizability parameter $\alpha_p$.
Note however that only fine tuning was needed since in the end
the values of the $\alpha_p$ for different configurations turned to 
be very close to each other (see Table~\ref{sets}).

\begin{figure}
\centering
\epsfig{figure=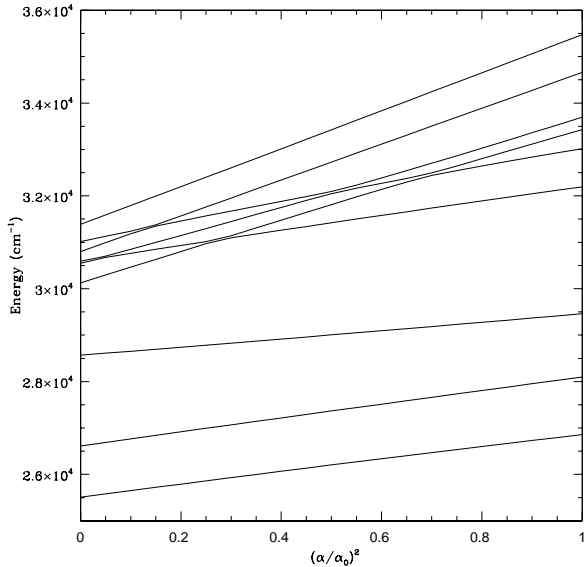,scale=0.4}
\caption{Odd-parity energy levels of Fe~I with total momentum $J=3$ as
functions of the fine structure constant.}
\label{f3}
\end{figure}

\begin{figure}
\centering
\epsfig{figure=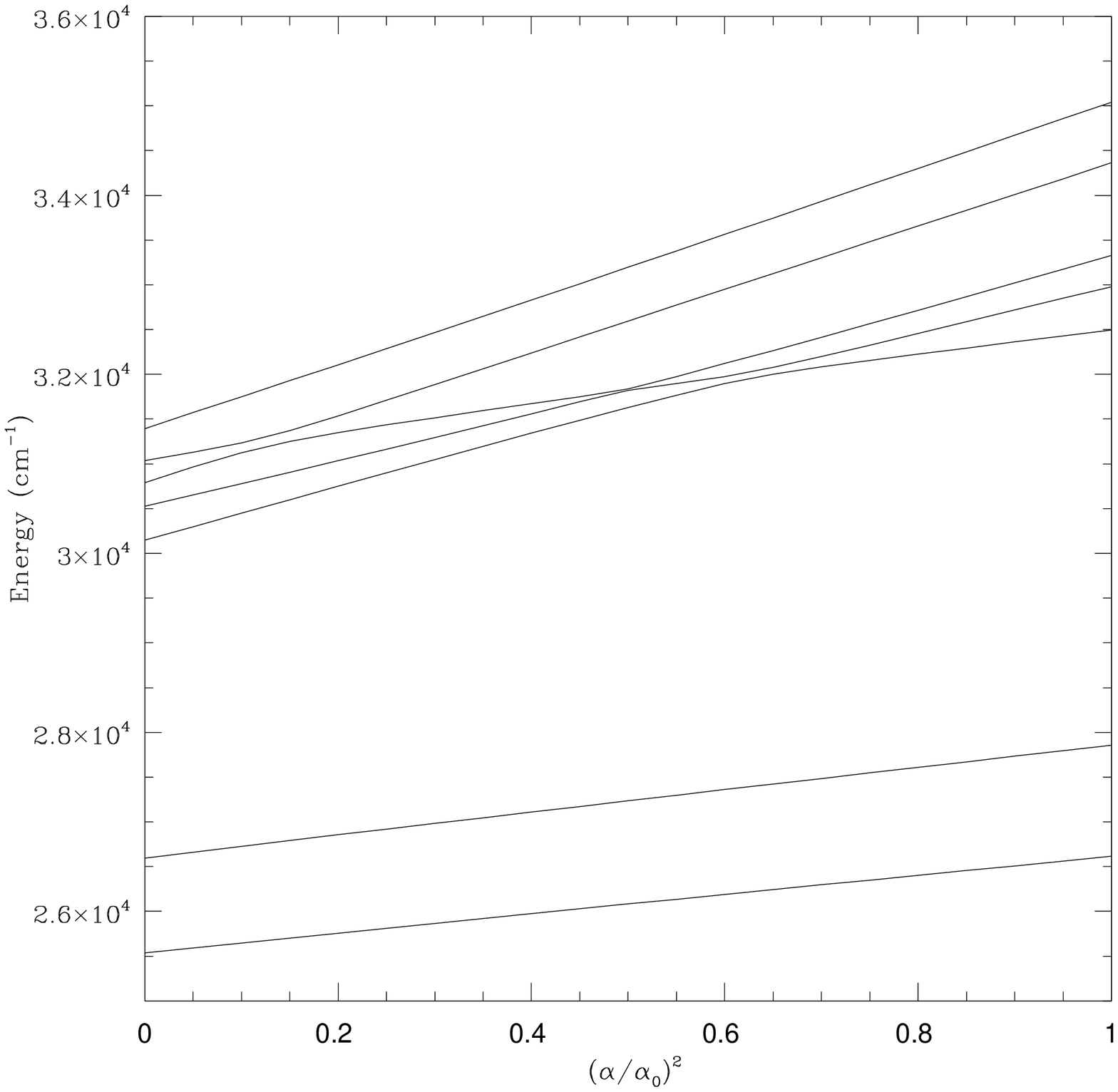,scale=0.4}
\caption{Odd-parity energy levels of Fe~I with total momentum $J=4$ as
functions of the fine structure constant.}
\label{f4}
\end{figure}

\begin{figure}
\centering
\epsfig{figure=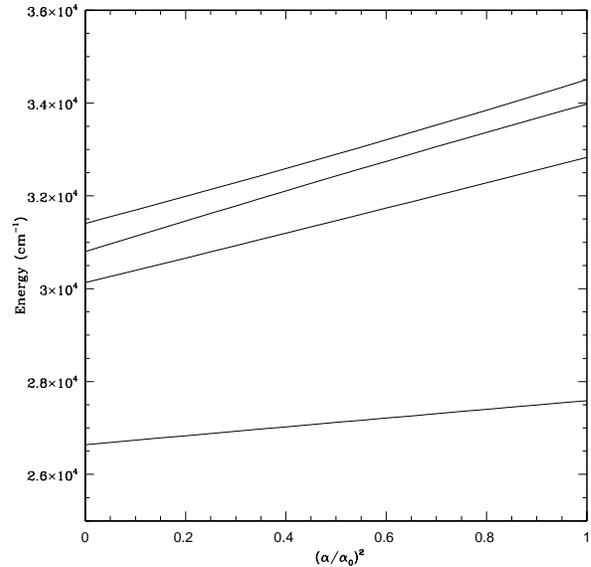,scale=0.4}
\caption{Odd-parity energy levels of Fe~I with total momentum $J=5$ as
functions of the fine structure constant.}
\label{f5}
\end{figure}

Another source of possible numerical instability of the results 
for particular states is level pseudo-crossing 
(see, e.g.~\cite{Dzuba01,Dzuba02}). Energies of the states when
considered as function of $\alpha^2$ may come close to each other
in the vicinity of the physical value of $\alpha$. Then small
error in the position of level crossing may lead to large
error in the $q$-coefficient which is actually the slop
of the curve $E(\alpha^2)$ (see, formula (\ref{qq}).
To investigate whether this is the case for Fe~I we plot the
energies of few low odd states of Fe~I with total momentum $J$ = 3, 4 and 5
as function of $\alpha^2$ from non-relativistic limit $\alpha=0$ to
the physical value of $\alpha$. The results are presented on 
Figs~\ref{f3}, \ref{f4} and \ref{f5}. As can be seen from the pictures,
there are multiple level crossing for states with $J=3$ and $J=4$.
However, all these crossings take place on safe distance from the 
physical value of $\alpha$ and are very unlikely to cause
the instability of the results. Another interesting thing to note
is that the energies are practically linear functions of $\alpha^2$ in
all cases.

\begin{table}
\caption{Energy levels (cm$^{-1}$) and $g$-factors of the lowest 
even states of Fe}
\label{Fe-even}
\begin{ruledtabular}
  \begin{tabular}{l l c r r r r}
Conf. & Term & $J$ & \multicolumn{2}{c}{Experiment\footnotemark[1]} &
 \multicolumn{2}{c}{Calculations} \\
 & & & \multicolumn{1}{c}{Energy} & \multicolumn{1}{c}{$g$} &
       \multicolumn{1}{c}{Energy} & \multicolumn{1}{c}{$g$} \\
\hline
$3d^64s^2$ & $a \ ^5$D & 4 &      0.000 & 1.500 &     0 &  1.4995 \\
           &           & 3 &    415.932 & 1.500 &   464 &  1.4997 \\
           &           & 2 &    704.004 & 1.500 &   790 &  1.4998 \\
           &           & 1 &    888.129 & 1.500 &  1000 &  1.4998 \\
           &           & 0 &    978.072 &       &  1103 &  0.0000 \\
           &           &   &            &       &	&         \\
$3d^74s$   & $a \ ^5$F & 5 &   6928.266 & 1.400 &  6862	&  1.3996 \\
           &           & 4 &   7376.760 & 1.350 &  7374 &  1.3496 \\
           &           & 3 &   7728.056 & 1.249 &  7779 &  1.2497 \\
           &           & 2 &   7985.780 & 0.999 &  8078 &  1.0000 \\
           &           & 1 &   8154.710 &-0.014 &  8275 &  0.0010 \\
           &           &   &            &       &	&         \\
$3d^74s$   & $a \ ^3$F & 4 &  11976.234 & 1.254 & 13040 &  1.2496 \\
           &           & 3 &  12560.930 & 1.086 & 13702 &  1.0835 \\
           &           & 2 &  12968.549 & 0.670 & 14171 &  0.6676 \\
\end{tabular}
\footnotetext[1]{NIST, Ref.~\cite{NIST}}
\end{ruledtabular}
\end{table}

Table~\ref{Fe-even} presents experimental and theoretical energies
and $g$-factors of the lowest even states of Fe~I. The $g$-factors are
useful for the identification of the states and for control of
configuration mixing~\cite{Dzuba02}. As can be seen the 
experimental data are reproduced in the calculations with very
good accuracy for both the $3d^64s^2$ and $3d^74s$ configurations.

\begin{table}
\caption{Energy levels (cm$^{-1}$), $g$-factors and relativistic
energy shifts ($q$-factors, cm$^{-1}$) for the states of 
of the $3d^64s4p$ configuration of Fe.}
\label{Fe-odd1}
\begin{ruledtabular}
  \begin{tabular}{l c r r r r r}
Term & $J$ & \multicolumn{2}{c}{Experiment\footnotemark[1]} &
 \multicolumn{3}{c}{Calculations} \\
& & \multicolumn{1}{c}{Energy} & \multicolumn{1}{c}{$g$} &
       \multicolumn{1}{c}{Energy} & \multicolumn{1}{c}{$g$} & 
 \multicolumn{1}{c}{$q$} \\
\hline
z $^7$D$^o$ & 5 & 19350.892~ & 1.597 & 19166 &  1.5987 &   490\\
            & 4 & 19562.440~ & 1.642 & 19390 &  1.6490 &   662\\
            & 3 & 19757.033~ & 1.746 & 19611 &  1.7485 &   891\\
            & 2 & 19912.494~ & 2.008 & 19793 &  1.9976 &  1092\\
            & 1 & 20019.635~ & 2.999 & 19921 &  2.9950 &  1237\\
            &   &            &       &       &	    &      \\
z $^7$F$^o$ & 6 & 22650.421~ & 1.498 & 21663 &  1.4997 &   582 \\
            & 5 & 22845.868~ & 1.498 & 21891 &  1.5005 &   827 \\
            & 4 & 22996.676~ & 1.493 & 22062 &  1.5026 &   982 \\
            & 3 & 23110.937~ & 1.513 & 22189 &  1.5029 &  1103 \\
            & 2 & 23192.497~ & 1.504 & 22282 &  1.5026 &  1184 \\
            & 1 & 23244.834~ & 1.549 & 22338 &  1.5029 &  1227 \\
            & 0 & 23270.374~ &       & 22366 &  0.0000 &  1246 \\
            &   &            &       &	  &	    &      \\
z $^7$P$^o$ & 4 & 23711.457~ & 1.747 & 22543 &  1.7470 &  491 \\
            & 3 & 24180.864~ & 1.908 & 23034 &  1.9136 &  983 \\
            & 2 & 24506.919~ & 2.333 & 23440 &  2.3309 & 1316 \\
            &   &           &       &	  &	    &      \\
z $^5$D$^o$ & 4 & 25899.987\footnotemark[2] 
                             & 1.502 & 26428 &  1.4979 &   999 \\
            & 3 & 26140.177~ & 1.500 & 26679 &  1.4984 &  1223 \\
            & 2 & 26339.691~ & 1.503 & 26924 &  1.4976 &  1450 \\
            & 1 & 26479.376~ & 1.495 & 27094 &  1.4971 &  1616 \\
            & 0 & 26550.476~ &       & 27174 &  0.0000 &  1705 \\
            &   &           &       &	  &	    &      \\
z $^5$F$^o$ & 5 & 26874.549\footnotemark[2]
                             & 1.399 & 27432 &  1.3999 &  880 \\
            & 4 & 27166.819~ & 1.355 & 27702 &  1.3517 & 1180 \\
            & 3 & 27394.688~ & 1.250 & 27947 &  1.2530 & 1402 \\
            & 2 & 27559.581~ & 1.004 & 28119 &  1.0041 & 1568 \\
            & 1 & 27666.346~ &-0.012 & 28213 &  0.0062 & 1680 \\
            &   &           &       &	  &	    &      \\
z $^5$P$^o$ & 3 & 29056.321\footnotemark[2]
                             & 1.657 & 29340 &  1.6643 &  859 \\
            & 2 & 29469.020~ & 1.835 & 29795 &  1.8307 & 1310 \\
            & 1 & 29732.733~ & 2.487 & 30118 &  2.4966 & 1594 \\
            &   &            &       &       &	     &      \\
z $^3$F$^o$ & 4 & 31307.243~ & 1.250 & 32356 &  1.2504 & 1267 \\
            & 3 & 31805.067~ & 1.086 & 32883 &  1.0885 & 1808 \\
            & 2 & 32133.986~ & 0.682 & 33263 &  0.6767 & 2177 \\
            &   &            &       &       &	     &      \\
z $^3$D$^o$ & 3 & 31322.611~ & 1.321 & 32032 &  1.3314 & 1456 \\
            & 2 & 31686.346~ & 1.168 & 32464 &  1.1662 & 1843 \\
            & 1 & 31937.316~ & 0.513 & 32750 &  0.5035 & 2119 \\
\end{tabular}
\footnotetext[1]{NIST, Ref.~\cite{NIST}}
\footnotetext[2]{States observed in quasar absorption spectra}
\end{ruledtabular}
\end{table}

Table~\ref{Fe-odd1} presents experimental and theoretical energies
and $g$-factors of the lowest odd states of Fe~I in which the
$3d^64s4p$ configuration dominates. Theoretical relativistic
frequency shifts ($q$-coefficients) are also presented. The 
$q$-coefficients were obtained by numerical differentiation 
using formula~(\ref{deriv05}). Note that only states with
$J$=3,4 and 5, for which electric dipole transition to the ground
state is possible are needed for the analysis. However, we
present $q$-coefficients for all states for better illustration of
the accuracy of the calculations. In the linear in $\alpha^2$
approximation the difference in $q$-coefficients for states of the 
same fine-structure multiplet is equal to the fine structure
interval between this states. As can be seen from Figs.~\ref{f3},\ref{f4}
and \ref{f5} the dependence of the energies on $\alpha^2$ is very
close to linear indeed. Therefore, comparing the data for the fine
structure and $q$ is another test of the calculations.

\begin{table}
\caption{Energy levels (cm$^{-1}$), $g$-factors and relativistic
energy shifts ($q$-factors, cm$^{-1}$) for the states of the
$3d^74p$ and $3d^54s^24p$ configurations of Fe}
\label{Fe-odd2}
\begin{ruledtabular}
  \begin{tabular}{l l c r r r r r}
Conf. & Term & $J$ & \multicolumn{2}{c}{Experiment\footnotemark[1]} &
 \multicolumn{3}{c}{Calculations} \\
 & & & \multicolumn{1}{c}{Energy} & \multicolumn{1}{c}{$g$} &
       \multicolumn{1}{c}{Energy} & \multicolumn{1}{c}{$g$} & 
 \multicolumn{1}{c}{$q$} \\
\hline
$3d^74p$   & y $^5$D$^o$ & 4 & 33095.937\footnotemark[2]
                                          & 1.496 & 32680 &  1.4511 & 2494 \\
           &             & 3 & 33507.120\footnotemark[2]
                                          & 1.492 & 33134 &  1.3492 & 3019 \\
           &             & 2 & 33801.567~ & 1.495 & 33466 &  1.1053 & 3423 \\
           &             & 1 & 34017.098~ & 1.492 & 33705 &  0.1658 & 3754 \\
           &             & 0 & 34121.58~~ &       & 34007 &  0.0000 & 3723 \\
           &             &   &           &       &	  &	    &      \\
$3d^74p$   & y $^5$F$^o$ & 5 & 33695.394\footnotemark[2]
                                          & 1.417 & 32522 &  1.3964 & 2672 \\
           &             & 4 & 34039.513~ & 1.344 & 33029 &  1.3913 & 3021 \\
           &             & 3 & 34328.749~ & 1.244 & 33404 &  1.3881 & 3317 \\
           &             & 2 & 34547.206~ & 0.998 & 33705 &  1.3777 & 3536 \\
           &             & 1 & 34692.144~ &-0.016 & 33909 &  1.3347 & 3678 \\
           &	         &   &		  &	  &	  &	    &      \\
$3d^74p$   & z $^5$G$^o$ & 5 & 34782.416~ & 1.218 & 33978 &  1.2487 & 3024 \\
           &             & 6 & 34843.94~~ & 1.332 & 33687 &  1.3330 &      \\
           &             & 4 & 35257.319~ & 1.103 & 34363 &  1.1406 & 3520 \\
           &             & 3 & 35611.619~ & 0.887 & 34661 &  0.9153 & 3864 \\
           &             & 2 & 35856.400~ & 0.335 & 34883 &  0.3505 & 3464 \\
           &             &   &           &       &	  &	    &      \\
$3d^74p$   & z $^3$G$^o$ & 5 & 35379.206~ & 1.248 & 34506 &  1.2209 & 3340 \\
           &             & 4 & 35767.561~ & 1.100 & 35042 &  1.0731 & 3697 \\
           &             & 3 & 36079.366~ & 0.791 & 35474 &  0.7671 & 4096 \\
           &             &   &           &       &	  &	    &      \\
$3d^74p$   & y $^3$F$^o$ & 4 & 36686.164~ & 1.246 & 35697 &  1.2425 & 3085 \\
           &             & 3 & 37162.740~ & 1.086 & 36227 &  1.0863 & 3487 \\
           &             & 2 & 37521.157~ & 0.688 &	  &	    &      \\
$3d^54s^24p$ & y $^7$P$^o$ & 2 & 40052.030~ & 2.340 & 40529 & 2.3278 &   \\
             &             & 3 & 40207.086~ & 1.908 & 40677 & 1.8883 & -2472 \\
             &             & 4 & 40421.85~~ & 1.75~ & 40926 & 1.7491 & -2287 \\
\end{tabular}
\footnotetext[1]{NIST, Ref.~\cite{NIST}}
\footnotetext[2]{States observed in quasar absorption spectra}
\end{ruledtabular}
\end{table}

Table~\ref{Fe-odd2} presents the data similar to those of Table~\ref{Fe-odd1}
but for the states where the $3d^74p$ and $3d^54s^24p$ configurations
dominate. The values of the $q$-coefficients for the states of the 
$3d^74p$ configuration are larger than those of the $3d^64s4p$ configuration.
This is due to additional contribution from the $4s - 3d$ singe-electron
transition as it was explained above.

It is interesting that similar to the case of the ion Fe~II~\cite{Dzuba02}
neutral iron also has some negative shifters($q<0$). Corresponding states
belong to the $3d^54s^24p$ configuration. Negative sign of $q$ is due to
the dominant contribution from the $4p -3d$ single-electron transition.
The data are presented in Table~\ref{Fe-odd2}. Note however that the spin
of these states is different from the spin in the ground state. This means
that the electric dipole transition is suppressed by conservation of spin
and goes only due to relativistic effects. This in turn probably means
that the transitions may be too weak to be observed.  

We estimate the accuracy of present calculations of the $q$-coefficients 
to be on the level of 20 to 30\%. The results were obtained with a very
simple method which uses small number of basis functions and some
semi-empirical fitting. The main challenges for more accurate
calculations are strong configuration mixing and large number of
valence electrons.  
Further development of the methods or the use of supercomputers 
might be needed for better accuracy of the calculations.

\section{Conclusion}

We have calculated relativistic frequency shifts for a number of the
lower odd states of Fe~I. Some of these states were observed in the 
quasar absorption spectra. Calculations show that due to strong 
configuration mixing the values of the shifts are large and vary
significantly between the states. This makes Fe~I to be a good
candidate for the search of variation of the fine structure
constant in quasar absorption spectra.

\section*{Acknowledgments}

We are grateful to Prof. P. Molaro for brining to our attention
lines of Fe observed in quasar absorption spectra.
The work was funded in part by the Australian Research Council.

\end{document}